# Artificial Intelligence Paradigm for Customer Experience Management in Next-Generation Networks: Challenges and Perspectives

Haris Gacanin and Mark Wagner

Nokia Bell Labs, {name.surname}@nokia-bell-labs.com

*Abstract* – With advancements of next-generation programmable networks a traditional rule-based decision-making may not be able to adapt effectively to changing network and customer requirements and provide optimal customer experience. Customer experience management (CEM) components and implementation challenges with respect to operator, network, and business requirements must be understood to meet required demands. This paper gives an overview of CEM components and their design challenges. We elaborate on data analytics and artificial intelligence driven CEM and their functional differences. This overview provides a path toward autonomous CEM framework in next-generation networks and sets the groundwork for future enhancements.

*Keywords* – Artificial intelligence, network optimization, network management.

## I. Introduction

The fifth generation (5G) networks are expected to support dynamic networking profiles with automated service provisioning [1], [2]. This sets the stage for challenging network requirements, where demands on resources are often dynamic and unpredictable depending on the content. Typically, issues that arise have been handled through an operator help desk. These are often time consuming and can result in negative customer feedback. This has led operators formulating strategies to address issues without requiring actions by the customer [3]. It has also given rise to customer experience management (CEM), which is essentially the practice of overseeing and responding to customers interactions with a network to improve the customer experience (CXS). Quantifying CXS requires understanding the combined behavior of humans and the network to evaluate the quality-of-experience (QoE) [4]. CEM builds intelligence across different functional layers by recognizing how customer service requests impact resources down to radio access [3] – [5].

Recently, a proactive and self-care customer management functions supported by data analytics (DA) have given insight into understanding CXS indirectly [3], [6], [7]. Related works on data-driven optimization have led the groundwork for end-to-end (E2E) network intelligence, allowing operators to consider different data types and sources to derive QoE models based on machine learning (ML) [6]. Such approaches unveil hidden patterns and correlations necessary that may not be directly visible to business operations [3].

The diverse nature of devices and services in 5G makes it difficult for traditional rule-based decision-making to adapt effectively to changing network conditions and provide optimal CXS [8]. Hence, CEM is becoming increasingly relevant in creating direct relations between desired CXS and an operator's decisions. This article provides an overview of state-of-the-art CEM requirements and future challenges facing stakeholders. Ultimately, we discuss an artificial intelligence (AI) driven concept in the context of a broader position and vision of autonomous CEM. The growing importance of proper understanding of complex business and network behavior is stressed to unveil necessary future research directions.

## II. CEM Components

CEM strives to understand the experience and interactions a customer has with a network. Ideally, CEM would identify the root causes of a problem without relying directly on interactions with customers. With proper sensor design, the metrics necessary to derive actionable insights are network quality-of-service (QoS) and customer QoE [9]. In our case, a sensor is either a physical device or any arbitrary way to measure certain metrics in the network.

### A. Network QoS

Network QoS deals with objective and technical metrics such as throughput, round-trip delay, packet loss, jitter, etc. at the network level. This differs from QoE, which has metrics such as frame rate, resolution, etc. and take place at the application level. Regardless of how QoS is implemented, it incorporates (*i*) non-technical requirements of the customer expressed through a service level agreement (SLA); (*ii*) E2E network connectivity, (*iii*) service observations presented in a technical fashion by the operator; (*iv*) and a report on the service experienced by the customer [4].

QoS cannot capture an E2E customer experience because its definition relates to network service and resource availability, not CXS. Often there is a significant difference between QoS at the network level, and what the customer experiences as quality at the application level [9]. To give a few examples, in voice and video use cases, QoS monitoring can be designed to track packet loss, delay, throughput, etc., depending on the target service. However, for data and resource management the relevant parameters may be at radio access, such as the received signal strength (RSS), error rate, etc. Smart factory applications utilize the network to provide real-time service information (e.g. manufacturing status) on any issue that may arise. Consequently, network issues causing delay of service information or diagnostics can generate high costs in real-time. In a smart city/building/grid scenario, relevant parameters are often time sensitive and may come from various sensors such as temperature and humidity. Thus, the achieved network QoS may not guarantee or describe an expected customer QoE.

### B. Customer QoE

Unlike QoS, where an objective and technical metric is provided, QoE is defined as the customer's subjective perception of delivered application. QoE is often measured by the mean opinion score (MOS) for web, audio and video applications, but its usefulness is often brought into question [4]. QoE includes the effects of E2E system elements ranging from interfaces, terminals, network and content with metric examples such as jerkiness, frozen frames, blurriness, etc. QoE can be divided into subjective elements such as experience, emotions, expectations, and objective elements consisting of both technical and non-technical aspects of services [4], [10]. In principle, technical aspects are throughput, connectivity, network/service coverage, and device functionality, while service provisioning, pricing and service content are some examples of non-technical aspects.

In general, QoE is a non-parametric, nonlinear function of QoS where, if network optimization is based solely on QoS, it may not lead to optimal QoE [9]. Defining an analytical relationship between technical aspects of QoS influencing QoE is not straightforward, if not impossible. This is because QoE is not directly captured by network measurements and random nature of human behavior. For example, video services with high bandwidth demands and moderate network QoS may be significantly impacted at the QoE level. This is mostly due to unpredictable customer demands and network neighborhoods that at certain moments may add additional interference, further degrading network QoS stability. Thus, QoE alone is not suitable for network root-cause analysis due to a lack of analytical models capturing relationships from application down to radio access. Therefore, data-driven modeling becomes very relevant as a future innovation enabler.

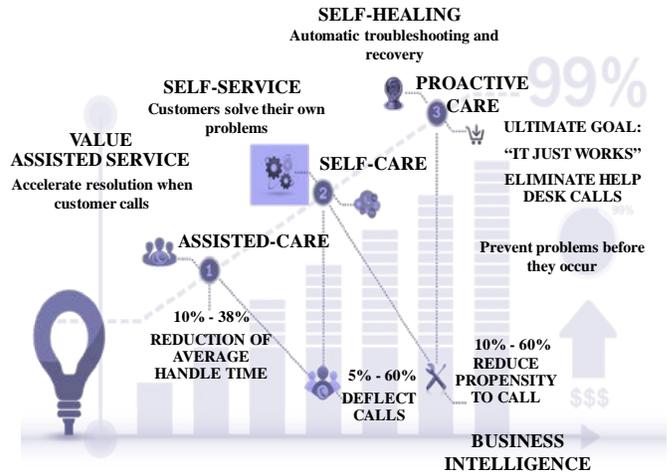

Figure 1. How improvements in business intelligence shift from assisted-care and self-care to proactive-care CEM strategy.

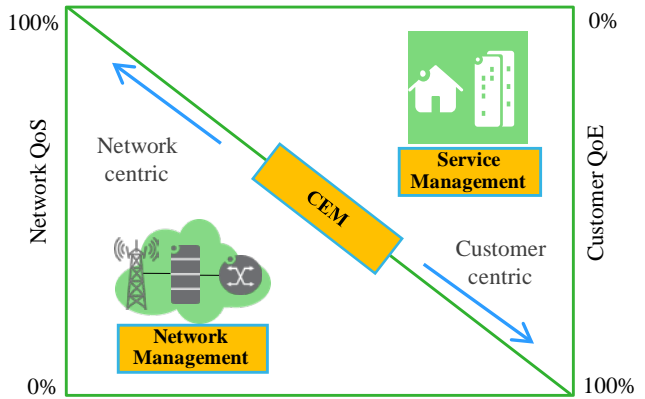

Figure 2. Network, service and customer experience management relationship: network centric management with 100% support of Network QoS and customer centric management with 100% support of QoE.

## III. CEM CHALLENGES IN 5G

### A. Operator-related Challenges

In 5G, fulfilling customer demands creates many new challenges for the operator to address. For example, over-the-top (OTT) services that increase network traffic per node may lead to significant revenue drains. Operators have little influence as to when, where or how-long customers place these demands on the network. In this case, operators utilize DA to predict and profile customers' usage patterns for planning and optimization [11].

To this end, self-care and proactive-care systems were introduced, moving away from assisted-care and towards more autonomous CEM with advancement in business intelligence, as illustrated in Figure 1. The shift away from assisted-care CEM is driven by reduction of operational costs and focused on automated recovery actions [3]. Self-care systems aim at data-heavy applications such as augmented virtual assistants, context-aware network optimization, customer-driven planning, and autonomous deployment [12]. The major challenge in this case is coupling customer demands and network resources in real time. This may be accomplished by utilization of supervised, unsupervised, and reinforcement learning with methods mentioned in Table 1 [3], [6], [9], [12].

Multi-tenant network capabilities should allow several service operators to share network physical infrastructure with tenants ranging from mobile network operators and OTT service providers, to vertical industries. This leads to challenges in optimizing network resources needed by the applications (i.e. network slices in 5G) [1]. It requires CEM frameworks to be designed with modular functions (sensing, perception, learning, etc.) across different and independent operational levels (e.g. engineering, marketing, etc.).

### B. Network-related Challenges

Figure 2 illustrates the multi-dimensional relationship between NM, service management (SM) and CEM. In this figure, as we move along the CEM line we transit between the network-centric approach in NM, where QoS is fully utilized (100%) and QoE is underutilized (or not utilized at all), toward customer-centric SM where QoE is fully utilized (100%) and QoS is underutilized. The former is common for network operators, while the latter is mostly used by OTT providers. The optimal point would strike a balance depending on the service and customer demands, building a flexible and context-aware network architecture.

5G networks consist of heterogeneous hardware resources that utilize general-purpose and dedicated hardware to host a wide variety of network functions. Development of normalized application programmable interfaces (APIs) for operations, control and management is required to enable customer-centric orchestration of services that allow flexibility in network design to customize service delivery [3].

Today's networks are designed to deliver specific services by gathering users around it – a network-centric approach. In 5G, the network is designed and built to adapt to customer demands (i.e customer-centric approach) rather than expecting customers to adapt to the network as it is [1]. This leads to the application of DA and ML in network operations. Without intervention from an operator's help desk, CEM needs to adaptively perceive and analyze a network state to cope with faults in real-time or before they occur (i.e. proactive analytics). These may include network service outages, deployment planning, and inefficient resource utilization. Even today, relationships between state analysis and resolution actions (recommendations) are highly complex and cannot be captured by human-defined rule-based logic [6], [9], [11].

Creation of intelligent agents requires experimentation with different AI/ML techniques (i.e. non-linear and probabilistic reasoning methods) such as belief networks, Markov models, neural networks, reinforcement learning, etc. For example, network planning and deployment in ultra-dense scenarios would leverage AI to enrich the optimization process, while multi-agent learning and customer-guided decision making remain open questions in location-search for deployment of new access points (APs) [12]. Further examples include network utilization through autonomous resource allocation, real-time optimization based on customer demand, load prediction, coverage optimization, etc. In these instances, AI/ML would be utilized in mapping network precepts to present or past actions.

Today, operators are collecting large volumes of data for analytics [3], [6]. The data collection should be dynamically controlled and adapted per use case in the statistically correct volumes to limit too much or too little data. Attention needs to be taken with assumptions that use cases are highly correlated (e.g. coverage, cell ID and load balancing), but evaluated independently. Finally, data-driven CEM is highly dependent on standardization and normalization of multi-vendor data-models [11]. A common northbound interface from the operator to the network and a southbound interface that controls the network resources should be supported by multi-vendor equipment. For example, coordination with respect to service mobility requirements cannot be supported by a single mobility function in a 5G network [1]. Hence, the mobility function must follow a modular and adaptive framework to modify network configuration according to service requirements. For customer location-awareness, we also need to carefully select network and customer data sets in the time domain to avoid solving problems in one area, just to relocate them in another.

### C. Customer-related Challenges

Network-centric management (QoS) is largely dominating today due to more tangible and well-defined functions. A shift to customer-centric management is limited in its ability to define accurate QoS-to-QoE models down to radio access as they become far too complex to be analytically tractable. There are several ways to perform QoS to QoE mapping [9]. In the case of ML, the mapping model is generated with the aim of inferring the QoE of an end user. Watching a video for example, the experimentally obtained data are split into training and validation sets. The training set is used to learn different models with the aid of ML algorithms such as random forest regression. The accuracy of learned models is then calculated on a validation set and models are used to predict QoE based on newly observed QoS inputs.

While implementing these frameworks, it's important to avoid intrusive data collection mechanism that interrupts service delivery, which are unacceptable from the perspective of customer SLAs. CEM requires efforts across all communication layers and network domains, where different functions such as network access selection, resource allocation, QoS mapping, session establishment, and source coding need to be adaptable to the customer's QoE. The critical issue is adjusting the QoE metric at the AP based on customer service and behavior (e.g. emotions, motion, etc.) for various devices and locations. Therefore, the shift from QoS-centric to QoE-centric networks is still emerging and an open research question [5].

## IV. STATE-OF-THE-ART CEM

In recent years, the primary limitation of CEM has been its independent management of QoS and QoE. This significantly limits CEM's impact on network and business performance and necessitates the need for a more integrated approach. Today, CEM introduced with DA gets data sources for modeling and knowledge discovery using the business (i.e. marketing, strategy and care) and network (i.e. engineering, planning and deployments) operation levels [3], [6]. At the business operation level, DA is used to create insights into customer behavior primarily through non-technical QoE indicators. For example, a help desk support where customer tickets are correlated with network alerts using supervised learning techniques to perform ticket classification. At the network operation level, DA performs the analysis using the radio, medium access and network protocols. Here, network parameters are utilized for diagnostics and troubleshooting of service quality in real-time or offline [9]. In another example, internet protocol traffic management [7] is applied to (*i*) prioritize certain types of traffic at busy times or areas; (*ii*) slow traffic that's not time critical; (*iii*) maintain customer SLAs by limiting traffic to the heaviest users; and (*iv*) sustain service quality for specific content. However, due to a lack of understanding of network dynamics, just using the business and network operation levels for DA is often inadequate to understand the root cause of customer problems.

### A. Integrating QoS and QoE

Many QoE and QoS management frameworks for 5G networks have been presented with flexible orchestration and coordination mechanisms. General E2E CEM frameworks for next-generation networks have been suggested [10], but implementation is highly challenging in multi-vendor and multi-operator environments. An integrated view of 4G networks has been introduced that considers network QoS optimization at communication layers, while directly monitoring QoE [13]. In this approach, it was advocated that the critical requirements for efficient QoS and QoE monitoring are dependent on AP control and performance parameters. A framework for QoE assessment in cellular networks with dynamic QoS management from the customer perspective has been presented [14]. The framework enables tracking capabilities of instantaneous changes of QoE. Design challenges of QoE management in cellular networks have also been presented [5]. The control, monitor and manager functions are presented having implementation challenges highly dependent on the domain expert and choice of QoE model. And it's not clear how integration with existing business systems would be possible. In general, metrics thus far are gathered and analyzed in isolation, and are unable to capture the full CXS life cycle.

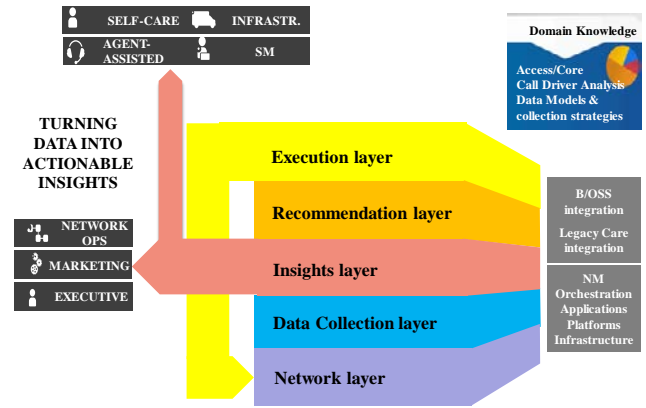

Figure 3. State of the art DA-driven CEM Framework with different functional layers and their relationships to operator systems.

### B. DA-driven Approach

With the advancement of data storage and large-scale distributed processing platforms (e.g. Hadoop), DA has far reaching potential to reveal network insights. It has allowed network operators to collect high volumes of information to be used for actionable insights [3], [6], [7]. DA models have been proposed to find patterns (hidden models) describing the relationship between network QoS and customer QoE [1], [15]. Thus, DA and graph-based inference are finding their way in management systems as a process of discovering information about the network status and suggesting proactive recommendations.

In fact, a DA-driven CEM framework has been practically tested for traffic and network management [7]. The approach led to useful insights and improved network planning and operation. Some DA-driven CEM models have also targeted business objectives directly by proper selection of a QoS/QoE model to predict customer behavior and understand what influences customer relationships [3]. In these models, the diagnostics and troubleshooting of the network is based on Bayesian inference and gives insight into the complex relationships between customer – operator lifecycle and the operator's business tools. We see that careful design of DA-driven CEM enables efficient monitoring of network performance with domain expert problem identification and troubleshooting. It provides proactive planning and optimization to cope with new services, applications, devices, and dynamic policy controls, reducing operational and capital expenditure [3]. As well, we can gain business insights such as market trends, market penetration analysis, geo-analysis and peer network performance comparison.

Many new directions have arisen to investigate the impact of feature set expansion and customer awareness for QoE forecasting services. One of the most critical areas of research is the design of efficient and non-intrusive data interfaces. For instance, opt-in software extensions on customer devices that perform a QoE forecasting service adds another dimension to the customer data, leading to more accurate and comprehensive QoE models [11]. This would give operators much broader insight into the CXS and allow for more personalized customer care.

### C. DA-driven Framework

Figure 3 captures the conceptual design of state-of-the-art DA-driven CEM framework with key functional layers execution, recommendation, insight, network, and data collection. Through different interfaces, the framework interacts with operator's segments such as services, operations, engineering, executive, customer care and marketing. This allows the framework to integrate with diverse data sources (e.g. customer devices, cloud services, radio access, network switches, etc.) that include dynamic and flexible southbound interfaces to and from the data collection layer.

The data collection layer interfaces with the network and applications to gather measurements pertinent to network QoS and customer QoE. Domain experts then utilize ML techniques to describe insights from these metrics.

The insights layer utilizes the necessary QoS-to-QoE (CEM) models by using DA-driven analysis. Since data collection and analysis is mostly done offline, the domain experts require enough knowledge to monitor network statistics, identify trends, generate efficient DA models, and provide new recommendation rules if needed. The insights are still dependent on domain experts and rule-based logic with mostly independent analysis at the operational levels.

The recommendation layer leverages human domain knowledge to provide insights and triggers rule-based actions at the service management layer [3]. Such actions are used to adjust network configuration, but, so far, DA-driven CEM does not support the root cause analysis due to lack of learning and reasoning at the insights layer.

The execution layer handles interfaces to network (e.g. SM/NM, traffic management) and customer functions (e.g. care system) and relies on cloud technology to create configuration adjustments at different network locations. Thus, the execution layer processes and turns data into actionable insights by means of DA and domain knowledge [3].

To date, DA-driven CEM is mostly applied to higher-layer protocols to inspect traffic patterns or customer and service modelling [3], [7]. This is primarily driven by the fact that DA at radio access is not properly understood and consequently largely neglected [11]. Intelligence of DA-driven CEM frameworks are limited by rule-based analytics and "if-then-else" recommendation and execution logic, where the accuracy of decisions depends on the ability of experts at different operational levels to understand and predict the E2E network performance. The shift to real-time frameworks is highly customer and service dependent and requires more integration among different CEM layers [1], [3].

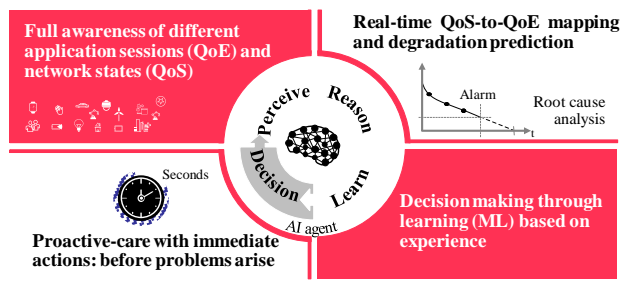

Figure 4. Conceptual functions and enablers of autonomous (AI-driven) CEM targeting proactive-care and problem root cause analysis.

## V. TOWARDS AN AI-DRIVEN CEM: COMPONENTS AND CHALLENGES

Figure 4 illustrates the conceptual functions and enablers of autonomous CEM. The intelligent agent needs to be fully aware of different network and application functions [1]. From the agent's perspective, the sessions are fully observable and customizable environments (e.g. utility-based, goal-based, etc.) [15]. The agent can facilitate a large range of devices and maintain intelligent traffic management and a flexible and adaptive response time function with proactive actions is likely required [1]. Lastly, the automated and data driven analysis needs to be enabled through learning. CEM framework layers described in Sect. IV should be observed as a logical grouping of AI functions such as sensing, perception, reasoning, decision making, optimization and learning [15].

The functional design of an AI-driven framework is illustrated in Figure 5, while some commonly used models and characteristics of each function are summarized in Table 1. The CEM agent can be implemented in different forms such as rule-based system, ontology-based system, and case-based reasoning using knowledge base (KB), among others [15]. For example, in the last instance a KB stores previous experiences in the form of a *problem, action and fitness* triplet. A *problem* is a vector referring to sensed measurements that perceive the current situation of the network QoS and customer QoE associated to an application (i.e. network slice). For each stored *problem*, an *action* can be performed which is essentially setting a new operational parameter in the network. After the *action* is applied, the *fitness* of this *action* is calculated based on the degree of customer QoE. For example, the *fitness* may be defined as a ratio between the achievable throughput and the demand which represents an estimate of the QoE.

### A. Data Collection Layer

The AI-driven framework will adopt an intelligent agent that is capable of sensing and perceiving the data at the network layer and map it to customer satisfaction. Sensing and perception functions help establish the data collection layer and interface with the network through programmable APIs. The APIs handle large amounts of dynamic, flexible, scalable, and enriched data with complementary information about each layer's behavior. These may include radio level measurements such as received signal strength, link level retransmissions or subscriber data plans and billing information at business/operation and support systems. Sensing the environment involves continuous collection of network measurements, application parameters and data from other external systems, such as radio-frequency identification and the Internet-of-Things. These measurements are then transformed to sensors by the perception module. For example, we can define a location sensor where the received signal level and round-trip delay can be used with application data to calculate customer location and optimize location-based services. The output of the data collection then triggers functions as illustrated in Figure 5.

### B. Insights Layer

The insights layer aims at the root cause of an application's unwanted behavior (e.g. rogue traffic). Insights trigger actions to re-configure the network (e.g. de-prioritizes, throttles, or removes this traffic) or deploy new services through policy-based decision-making. The layer relies and combines existing knowledge in KB with reasoning of network state in a probabilistic fashion. Table 1 summarizes some of the more relevant characteristics and models. For example, when the perceived network throughput is lower than the customer demand, the agent triggers the reasoning function to evaluate the currently perceived state with previously experienced situations in KB. Here, the KB builds and retains knowledge of the network performance and expected customer QoE after applying each action plan. The agent retrieves the most relevant case from the KB and reuses the corresponding *action* to resolve the current *problem*. This means that the current perceived issue is compared to all stored problems in the KB by means of a matching factor (e.g. distance metric). This is called deterministic reasoning, but other approaches based on probabilistic reasoning have been considered [15].

### C. Recommendation/Execution Layer

The recommendation and execution layers encompass an agents' decision-making, optimization and learning functions. These drive configuration changes at different network locations. Examples of the agent actions are service provisioning, setting operational network parameters, device association, deployment initialization, proactive SLA updates, etc. The agent stores previous *action* plans and evaluates their quality through direct or indirect customer feedback. The *action* and *fitness* of the matching case are evaluated in the decision-making function. The function determines whether the *action* of the most matching case can be reused or whether a new *action* needs to be re-calculated.

Decision-making functions checks both the distance metric and the *fitness* value of the retrieved case. If an observation is not true, then the KB holds and there's no matching case. In this instance, two scenarios are possible (*i*) a new case must be retained in KB or (*ii*) the best matching case has a suboptimal action that should be re-calculated. The optimization function would then calculate a new action to be executed and saved back to the KB.

The optimization search direction for new actions can be implemented through an exploitation and exploration algorithm [15]. *Exploitation* greedily optimizes the network metrics within a limited search space that appears to be promising, while *exploration* tries to discover new search spaces that can lead to more promising solutions than are currently available.

The learning function controls re-calculation and retains the entries in the KB by improving *fitness* accuracy of each action [12]. When the user is involved in providing direct or indirect feedback, semi-supervised learning is adopted. Strictly supervised learning techniques are not applicable when there is a lack of full knowledge about the environment (i.e. network environment, traffic demand, true customer satisfaction, etc.). In this scenario, reinforcement learning can be used where decision making and reasoning are combined [15].

Unlike the approaches in Sect. IV, the AI-driven approach shifts from model-based (i.e. DA) to model-free learning as the former requires a long time to build insights that are sensitive to the accuracy of observations and domain experts. Some implementation relevant characteristics and models of this function are listed in Table 1.

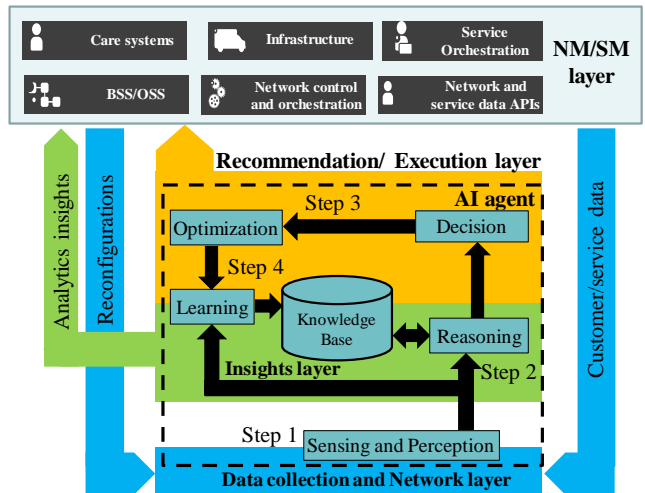

Figure 5. Functional diagram of AI-driven CEM framework: functions and dependencies: sensing/perception, reasoning, decision making, learning and optimization.

## VI. Future Directions

With these new DA approaches to CEM, new challenges arise and must be addressed. For instance, service orchestration in CEM is a critical element in designing new network slices. Policies in these areas need to introduce APIs to access physical and virtual network segments of multiple operators. Identification, sharing, and design of those interfaces will make it challenging to support many tenants with dynamic SLA delivery.

A clear understanding of the relationship between QoS and QoE through a comprehensive knowledge base at radio access will be essential. Thus, supporting root cause analysis of CXS, the radio access diagnosis and troubleshooting cannot be neglected. Without this aspect, we may not understand the network and customer behavior, nor properly design recommendation actions. To date, AI with ML has not been widely studied for proactive CEM and another potential research direction.

Data collection from multiple (non-standardized) sources, where two different vendors may have slightly different implementation of the same parameter, would require unification at the data collection layer. Aspects such as different data collection protocols, operational requirements to initiate collection sessions, support of mass vs. single data collection, and data collection frequency, all present challenges and will need to be considered in future CEM frameworks.

## VII. Conclusion

Different CEM framework designs have enabled network intelligence and key customer data to tease out critical business value drivers. Functional and implementation components enable the development of highly modular frameworks to identify insights and generate actions that optimize CXS. These frameworks support real-time, automated, and flexible APIs to facilitate integration of external systems such as OTT providers and customer service interfaces. This AI driven approach to CEM is key in meeting the technological demands of 5G and allow network operators and service providers to meet the ever-changing requirements of future experiences.

## Acknowledgments

The authors would like to thank Dr. Ramy Atawia for his valuable comments and advice. We also thank to anonymous reviewers for their suggestions and comments that helped to improve the manuscript.

| Functions | Characteristics | Models |
|---|---|---|
| Reasoning functions search for the best rational action in response to a state. An agent maintains a belief state that represents which states of the world are currently possible. From the belief state and a transition model, the agent can predict how the world might evolve in the next time step. Observables and a sensing function allow the agent to update the belief state. | • Deterministic - belief states are determined by logical formulas.<br>• Probabilistic - belief states are quantified as likely or unlikely | Boolean logic; Temporal models such as hidden Markov models, Kalman filters and dynamic Bayesian networks (special cases are hidden Markov models and Kalman filters); Multiple-object tracking by nearest-neighbor filter and Hungarian algorithm. |
| Decision-making: the maximization of expected utility in episodic or sequential decision problems. | • Logical agent cannot deal with uncertainty and conflicting goals;<br>• Goal-based agent deals with binary distinction between good (goal) and bad (non-goal) states<br>• Decision-theoretic (utility and probabilistic theory) agent can make decisions based on what it believes and what it wants with continuous measure of outcome quality. | Decision networks, decision-theoretic expert system, multiple agents (game theory), Markov decision process, dynamic Bayesian network. |
| Optimization (planning) is devising a plan of action to achieve goal. Optimization finds the best hypothesis within action space. | • Search-based problem-solving, Logical agent;<br>• Constrained optimization<br>• Heuristics | Hill climbing and simulated annealing for local search, convex optimization and linear programming for continuous spaces, backward (regression) and forward (progression) state-space search, graphs |
| Learning function is acquiring knowledge based on the observed states after applying the action. Learning improves reasoning by enriching the knowledge or experience used in reasoning function. | • Supervised learning with labeled data<br>• Unsupervised learning with unlabeled data<br>• Reinforcement learning based on maximizing a cumulative reward with taken actions. | Regression, K-nearest neighbor, support vector machine, Bayesian network; Clustering, principal component analysis; Markov decision process, Q-learning, game theory. |

Table 1. Intelligent agent: functions, characteristics and models.